\documentclass[letterpaper]{article}
\usepackage[T1]{fontenc}
\usepackage{comment}
\usepackage{float}
\usepackage{booktabs}
\usepackage{amsmath}
\usepackage{xr}
\usepackage{geometry}
\usepackage{setspace}
\usepackage{caption}
\usepackage{hyperref}
\usepackage[backend=biber, style=chem-acs, articletitle=true]{biblatex}
\usepackage{graphicx}
\usepackage{float}
\newfloat{scheme}{htbp}{los}
\floatname{scheme}{Scheme}
\floatname{chart}{Chart}
\newfloat{graph}{htbp}{loh}
\usepackage{chemformula} 
\usepackage[version = 4]{mhchem} 

\usepackage{authblk}
\author[1]{Nanditha Sunil Kumar$^*$}
\author[1]{Prerna Paliwal$^*$}
\author[1]{Mikhail Popov}
\affil[1]{Department of Chemistry, University of Basel, Klingelberstrasse 80, 4056 Basel, Switzerland}
\author[2]{Xueyi Li}
\author[2]{Anyang Li}
\affil[2]{Key Laboratory of Synthetic and Natural Functional Molecule of the Ministry of Education, College of Chemistry and Materials Science, Northwest University, 710127 Xi’an, China}
\author[3]{Hua Guo}
\affil[3]{Department of Chemistry and Chemical Biology, Center for Computational Chemistry, University of New Mexico, Albuquerque, New Mexico 87131, USA}
\author[1]{Stefan Willitsch}

\title{\bf Single-molecule cold chemistry: Isotopic branching ratios of the \ce{H2O+} + HD reaction at cryogenic temperatures}
\date{* These authors contributed equally. Email: liay@nwu.edu.cn (Anyang Li), hguo@unm.edu (Hua Guo), stefan.willitsch@unibas.ch (Stefan Willitsch).}

\begin{document}

\maketitle

\begin{abstract}
Isotopic branching in ion–molecule reactions influences interstellar deuterium fractionation, yet its dynamics remain poorly understood for many important systems at astrophysically relevant temperatures. Here, we investigate H- and D-atom transfer in the H$_2$O$^+$ + HD reaction using single sympathetically cooled water ions in a cryogenic radiofrequency trap. Nondestructive identification of the products from 441 individual reaction events at a mean collision energy of 7~K reveals an H$_3$O$^+$/H$_2$DO$^+$ product branching ratio of $1.50 \pm 0.15$ demonstrating a preferential H-atom transfer. The experiments are rationalized using quasi-classical-trajectory calculations which indicate a transition from direct dynamics at higher energies to a complex-mediated reaction under the present experimental conditions and point to an important role of intramolecular vibrational redistribution affecting the isotopic branching in this regime. The present results extend measurements into an astrochemically relevant energy regime and establish cryogenic single-ion chemistry as a new platform for resolving competing reaction pathways.
\end{abstract}

\newpage


Ion--neutral reactions are often barrierless and thus dominate the chemical evolution in low-temperature environments where temperatures can be as low as a few Kelvin such as the interstellar medium (ISM) \cite{snow08a, larsson12a}. In this context, the water cation \ce{H2O+} is a key intermediate in the chemistry leading to gas-phase water formation in the ISM \cite{kumar18a, tran18a}. Its reactions with hydrogen isotopologues serve as a sensitive probe of H/D transfer and the resulting deuterium fractionation under the cold conditions of interstellar clouds \cite{ard14, song17a, xu17a}. In particular, the reaction of \ce{H2O+} with deuterium hydride HD is of key interest as it can proceed through two energetically allowed product channels,
\begin{equation}
\label{eq:reactionHD}
\mathrm{H_2O^+ + HD \longrightarrow}
\begin{cases}
\mathrm{H_3O^+ + D}, \\
\mathrm{H_2DO^+ + H}.
\end{cases}
\end{equation}
Previous laboratory work on this reaction employed rovibrationally state-selected \ce{H2O+} ions to measure integral cross sections and product branching ratios for both the H-transfer and D-transfer channels over a collision-energy range $E_\mathrm{coll}=0.03$--10~eV \cite{ard14,xu17a,xu12a}. These experiments revealed a strong dependence of the branching ratios on the collision energy suggesting that the degree of deuterium fractionation reflects not only spatial variations in the HD abundance of the interstellar medium \cite{balashev19a}, but also the local temperature, which controls the balance between competing reaction pathways \cite{xu17a,song17a}. Accompanying quantum wave-packet and full-dimensional quasi-classical trajectory (QCT) calculations modeled the energy-dependence of the H/D-transfer rates and rationalized them in terms of the differences in the centrifugally corrected potentials for both reaction channels \cite{song17a}. However, the lowest collision energy explored in these studies ($0.03$~eV $\approx 350$~K) remained orders of magnitude above the regime of a few Kelvins prevailing in cold molecular clouds. Extrapolations to low temperatures where subtle quantum effects such as quantum tunnelling, zero-point energy and resonances can become decisive \cite{wild23a} may fail considering the pronounced energy dependence of this reaction. Thus far, to our knowledge the only low-temperature data was reported by Hugo \textit{et al.} \cite{hugo09a} who used it as an \textit{in situ} calibration method for determining the contamination of HD in an \ce{H2} sample. Branching fractions of 57\% and 43\% were reported for the \(\ce{H3O+ + D}\) and \(\ce{H2DO+ + H}\) product channels, respectively, determined in a temperature-variable 22-pole ion trap. Although their apparatus operated down to 13~K, the exact reaction temperature for this specific measurement was not reported \cite{hugo09a}.

Here, we introduce an experimental platform for studying reactions of single Coulomb-crystallized molecular ions trapped under cryogenic conditions in ultrahigh vacuum. Our approach is related to cryogenic Coulomb-crystal experiments reported recently \cite{wang26a}, but relies on trapping single ions instead of large ensembles in order to achieve a high degree of control over the reaction partners and to faithfully emulate conditions prevalent in the ISM. In the present work, we used this experimental method for the investigation of the low-temperature reaction of single \ce{H2O+} ions with HD molecules. The neutral HD molecules were cooled by repeated surface capture and desorption enabling collision energies as low as $E_\mathrm{coll}/k_\mathrm{B}=7$~K (0.6~meV). Our data confirm a pronounced isotope effect that persists down to the low temperatures of the present study. The experimental results were rationalized by QCT calculations which indicate a switch of reaction mechanism from direct to indirect at the experimentally relevant collision energies. The current results establish cryogenic single-ion chemistry as an attractive platform for exploring ion–molecule reactions under astrochemically relevant conditions and for benchmarking theoretical models for low-temperature chemical reactivity. 


\begin{figure}[htb]
  \centering
  \includegraphics[width=\linewidth]{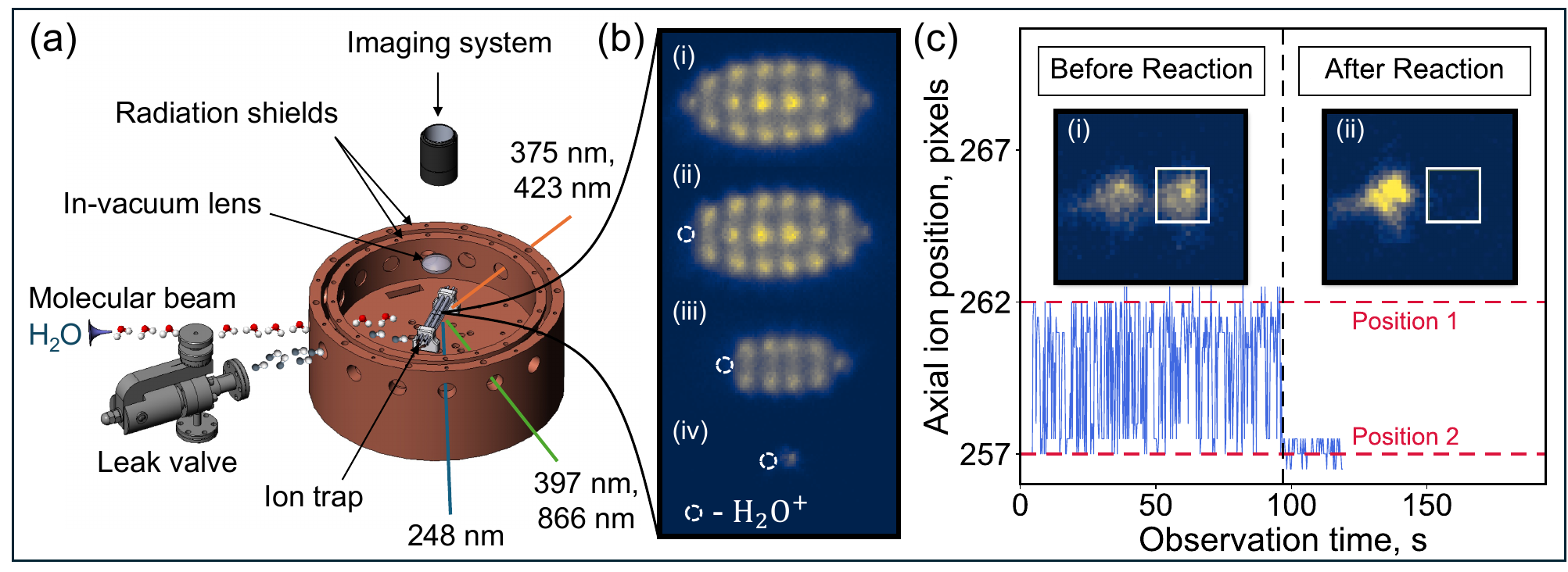}
\caption{\textbf{Generation, sympathetic cooling, and reactions of trapped single \ce{H2O+} ions.}
\textbf{(a)} Schematic illustrating the main components of the experimental setup. See text for details. \textbf{(b)} False-color fluorescence images showing the preparation of a two-ion \ce{Ca+}-\ce{H2O+} crystal: (i) a Coulomb crystal of laser-cooled \ce{Ca+} ions; (ii) the crystal after loading a single sympathetically cooled, nonfluorescing \ce{H2O+} ion (dashed circle); (iii) progressive size reduction of the crystal; and (iv) the resulting two-ion \ce{Ca+}-\ce{H2O+} string prior to reaction. 
\textbf{(c)} Representative experimental signature of the reaction obtained from monitoring the position of the fluorescence of the \ce{Ca+} ion along the longitudinal trap axis as a function of time. The blurring of the position of the \ce{Ca+} ion results from the repeated swapping of the positions of the two ions during resonant motional excitation. A discrete change in the fluorescence response at the point of reaction (vertical dashed line) signals a reaction of \ce{H2O+} with \ce{HD} heralded by the breakdown of the resonant excitation due to a change of the mass of the molecular ion.
The horizontal dashed lines indicate the two characteristic positions of \ce{Ca+} during motional excitation.
Inset: images of the \ce{Ca+} fluorescence (i) before reaction during resonant excitation, and (ii) after reaction with motional excitation suppressed.}
  \label{fig:crystal_and_tickling}
\end{figure}

The core of the present experimental setup is constituted by a linear-quadrupole radiofrequency (RF) ion trap \cite{willitsch12a} maintained at 5.4 K and surrounded by two cryogenic shields (Methods). Each experimental cycle was initiated with the preparation of a Coulomb crystal of several tens of laser-cooled \ce{Ca+} ions which served as a coolant for the water ion (Fig.~\ref{fig:crystal_and_tickling}(b)(i) and Methods). Single \ce{H2O+} ions were generated from a molecular beam of water by [2+1] resonance-enhanced multi-photon ionization (REMPI) inside the trap and their translational temperature was sympathetically cooled to a few millikelvin by interaction with the laser-cooled Ca$^+$ ions \cite{willitsch12a}. Molecular ions embedded in the crystal appeared as non-fluorescent (dark) spots within the Coulomb crystal (white dashed circle in Fig.~\ref{fig:crystal_and_tickling}(b)(ii)). Subsequently, calcium ions were selectively ejected from the trap by transiently adjusting the RF trapping potentials \cite{meir19a} to prepare a two-ion \ce{Ca+}--\ce{H2O+} string (Figs.~\ref{fig:crystal_and_tickling}(b)(iii) and (iv)).

To verify the chemical identity of the sympathetically cooled ions and track changes resulting from chemical reactions, the mass of the dark ion in each experiment was measured by a resonant excitation of the motion of the two ions in the trap using a time-varying electric field applied to one of the trap electrodes (Methods).  After verification of the initial presence of a water ion, the ions were continuously motionally excited at the frequency of the axial in-phase mode of the Ca$^+$-H$_2$O$^+$ two-ion crystal while being exposed to the co-reactant HD molecules (inset (i) in Fig.~\ref{fig:crystal_and_tickling}(c). A reaction event of the H$_2$O$^+$ ion was heralded by a disruption of the motional excitation at the original drive frequency due to a change of the mass of the sympathetically cooled ion (inset (ii) in Fig.~\ref{fig:crystal_and_tickling}(c). The identity of the product ion was subsequently established by resonant-excitation attempts at frequencies corresponding to the masses of the expected products. Under the present experimental conditions, the motional frequencies of the possible products which differ by a single mass unit were sufficiently different so as to enable their unambiguous identification (see Tab.~\ref{sec:frequencies} in the Supplementary Information (SI)). 

For studying the target reaction, neutral HD gas (Cambridge Isotope Laboratories Inc., isotopic purity $\sim$97\%) was dosed into the cryogenic environment using a leak valve. Before reaching the trap to react with the ions, the HD molecules had to diffuse through small apertures into the volume enclosed by the cryogenic shields of the vacuum chamber (Fig.~\ref{fig:crystal_and_tickling}(a)). Within the space enclosed by the shields, the molecules undergo multiple collisions, adsorption and desorption events with the cryogenic housings leading to thermalization and equilibration at the temperature of the walls. The partial pressure of HD in the reaction zone was determined to be $P_{\mathrm{HD}}=5.7(2.5)\times10^{-12}\,~\mathrm{mbar}$ (see Sec. \ref{sec:pressures} of the SI for details).

\begin{figure}[htbp]
  \centering
  \includegraphics[width=1\linewidth]{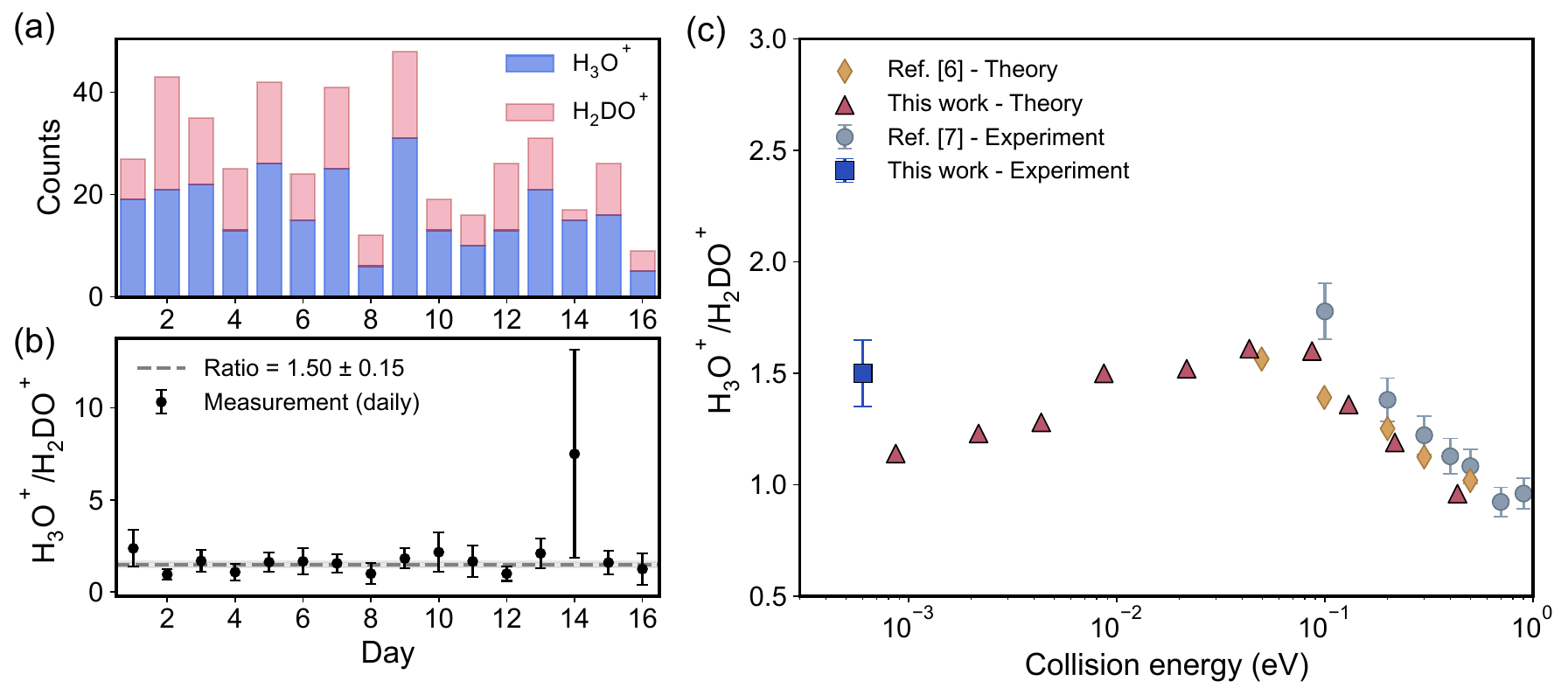}
\caption{{\bf Product branching in the \ce{H2O+ + HD} reaction.} \textbf{(a)} Counts of product ions of the title reaction obtained from single-ion reaction experiments on 16 different measurement days. \textbf{(b)} Measured branching fractions for formation of \ce{H3O+} (H transfer) and \ce{H2DO+} (D transfer) computed from the data in panel (a) for each measurement day. Error bars denote $1\sigma$ statistical uncertainties, assuming independent Poisson counting statistics. \textbf{(c)} H$_3$O$^+$/H$_2$DO$^+$ product branching ratio as a function of collision energy comparing the present result with quasi-classical trajectory (QCT) calculations and experimental data from the literature at higher energies. See text for details.} 
  \label{fig:branching_ratio}
\end{figure}

In the absence of the HD co-reactant, the trap lifetime of \ce{H2O+} was primarily constrained by its reactivity with \ce{H2} which typically constitutes the dominant residual gas in ultrahigh vacuum (UHV). Loss of H$_2$O$^+$ due to reactions $\mathrm{H_2O^+ + H_2 \longrightarrow H_3O^+ + H}$ therefore limited the achievable storage time $\tau_{\mathrm{bg}}$ of the ions and constituted the major background process affecting the study of the title reaction. The mean storage time $\tau_{\mathrm{bg}}$ of H$_2$O$^+$ limited by background reactions was subsequently determined from 20 independent reaction events under identical conditions. A result of $\tau_{\mathrm{bg}} = 4.5(1.0)~\mathrm{h}$ was obtained, from which a residual H$_2$ partial pressure $P_{\mathrm{H_2}}=3.0(0.7)\times10^{-14}\,~\mathrm{mbar}$ was determined (see Sec. \ref{sec:bgreactions} of the SI for details). This value is about two orders of magnitude lower than the partial pressure of the neutral reactant gas used in the the present experiments.

The present  [2+1] REMPI scheme has been shown to prepare \ce{H2O+} in its vibronic ground state \cite{uselman07a}. It can further be assumed that the population is initially spread over several low-lying rotational states dictated by rotational photoionization dynamics \cite{willitsch05b}. However, as the time-to-reaction is typically of order 100~s under the present conditions (see Fig.~\ref{fig:Lifetimes_H2_and_HD}b in the SI), the ions can at least partially rotationally thermalize by spontaneous emission and optical pumping by blackbody-radiation (BBR) in the cryogenic environment during that time. Since the trapping region is enclosed predominantly by the 5.4~K cryogenic shield, the BBR field experienced by the ions is dominated by this temperature, with only minor contributions from warmer surfaces visible to the ions through small apertures. The rotational temperatures can therefore be assumed to be close to thermal equilibrium: the $J_{K_aK_c}=0_{00}$ rotational state accounts for approximately $96\%$ of the population, whereas the $1_{11}$, $2_{02}$, and $2_{11}$ states contribute approximately $3.5\%$, $0.04\%$, and $0.01\%$. Here, $J$ denotes the rotational-angular-momentum quantum number and $K_a$ and $K_c$ the asymmetric-top pseudo quantum numbers. In summary, it can thus be assumed that the ions were translationally cooled to mK temperatures, prepared in their ground vibrational state and rotationally relaxed close to thermal populations.

Product branching ratios for the title reaction were determined by counting the number of \ce{H3O+} and \ce{H2DO+} products over 441 independent reaction events measured over 16 measurement days. The product mass and the time-to-reaction (defined as the interval between loading the dark \ce{H2O+} ion and the first observed reaction signature) were recorded for each event. Because the HD sample employed was only $\sim97\%$ isotopically enriched, this implied a $\sim3\%$ contribution from reactions with \ce{H2} present as major impurity in our HD gas sample. The measured product branching ratios were consequently corrected for the H$_2$ impurity (see Sec. \ref{sec:branalysis} in the SI for details). 

Figs.~\ref{fig:branching_ratio} (a) and (b) shows the statistics of the individual reaction outcomes as well as the H$_3$O$^+$/H$_2$DO$^+$ product branching ratios determined on each measurement day, respectively. The resulting impurity-corrected branching ratio, defined as the ratio of H-transfer to D-transfer product yields H$_3$O$^+$/H$_2$DO$^+$, was found to be $1.50(\pm0.15)$ averaged over all data. The quoted uncertainty represents the 1$\sigma$ statistical uncertainty assuming Poissonian statistics. 

To evaluate the effective collision energy associated with this result, it needs to be considered that the single \ce{H2O+} ions were sympathetically cooled to translational temperatures of a few millikelvin \cite{willitsch12a}, whereas the neutral HD molecules were assumed to thermalize with the cryogenic housing and therefore exhibit a Maxwell--Boltzmann velocity distribution at a temperature of $5.4$~K. Due to continuous RF excitation of the ions, a minor energy contribution from driven ion motion has to be considered as well. Overall, a mean collision energy $E_{\mathrm{col}}/k_B \simeq 7$~K (0.6~meV) was determined (see Sec.~\ref{sec:ecoll} of the SI for details). 

\begin{figure}[htbp]
  \centering
  \includegraphics[width=0.5\linewidth]{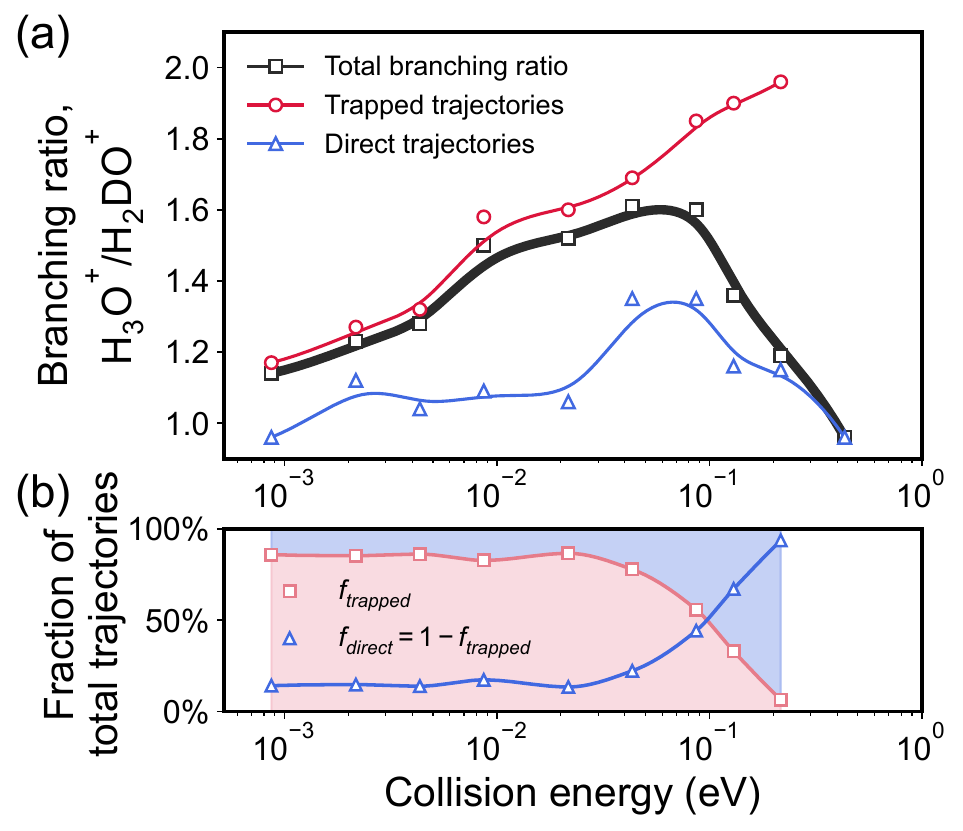}
\caption{{\bf Theoretical results from QCT calculations for the \ce{H2O+} + \ce{HD} reaction} (a) Branching ratio plotted as a function of collision energy for the different trajectories contributing to the reaction. (b) Fraction of trajectories $f$ corresponding to the direct and indirect ("trapped") mechanism as a function of the collision energy. See text for details.} 
  \label{fig:QCTresults}
\end{figure}

To rationalize the experimental observations, QCT calculations were performed to compute the branching ratio of the \ce{H2O+}+\ce{HD} reaction. The results within the collision-energy range from 1~meV to 1~eV are displayed together with the present and previous \cite{xu17a} experimental data in Fig.~\ref{fig:branching_ratio} (c), as well as with previous theoretical results from Ref.~\cite{song17a}. The present experimental value for the branching ratio of $1.50~\pm0.15$ at 0.6~meV is in qualitative agreement with the theoretical value of 1.15 at 0.9~meV. The QCT calculations predict that the branching ratio increases with collision energy, reaches a maximum around $40-80$~meV, and then declines again. This turnover is understood as the result of the superposition of two competing trends: a direct mechanism that proceeds without the reaction partners spending significant contact time, and a complex-forming (indirect) mechanism in which the reactants are temporarily captured as an ion-molecule complex in the pre-reaction well of the potential energy surface (see Fig.~1(a) of Ref.~\cite{song17a}) for a schematic) in which they can undergo extensive intramolecular vibrational energy redistribution (IVR). The competition of the two mechanisms is evidenced in Fig.~\ref{fig:QCTresults} which shows their contributions to the total branching ratio (panel (a)) as well as the fraction of trapped vs. direct trajectories (panel (b)) as a function of the collision energy. The indirect mechanism prevails at low energies, but gives way to direct dynamics in the energy region around the maximum in the branching ratio. 

The decline of the branching ratio with increasing collision energy beyond 100~meV has already been rationalized in Ref.~\cite{song17a}. There, it was concluded that the reaction proceeds via a direct mechanism in which the centrifugal barrier for the D-transfer channel is substantially higher than that for the H-transfer channel, thus favoring the production of H$_3$O$^+$. However, as the collision energy increases and thus more partial waves contribute, the difference between the two effective barriers diminishes, leading to a monotonic decline in the branching ratio. 

At the low collision energies at which the indirect mechanism prevails, the branching ratio obtained from the trapped trajectories decreases monotonically with decreasing $E_\text{col}$. The prolonged residence of the reaction partners in the pre-reaction well allows IVR to efficiently redistribute their initial translational energy among internal modes of the complex which seems to attenuate an isotopic preference of the reaction. The proposed role of IVR for the reaction is further substantiated by the correlation between the residence time in the pre-reaction well and the product branching ratio, as presented in Figure \ref{fig:complex_lifetime} of the SI. For trajectories with short lifetimes (i.e., those that escape the well quickly), the branching ratio is high. However, as the lifetime of the complex increases, the branching ratio decreases indicating that longer IVR periods lead to an increasing loss of the initial mass preference. 

The quantitative discrepancy between the present experimental and computed branching ratios at low energies may be due to the neglect of quantum effects in the theory, e.g., hydrogen tunneling, or inaccuracies in the potential energy surface. In addition, experimental factors such as isotope-dependent diffusion kinetics of the neutral reactant in the experimental chamber and fractionation induced by different condensation kinetics of HD and H$_2$ may play a role as well. The possible ramifications of such effects are discussed in Sec~\ref{sec:branalysis} of the SI.

In summary, the astrochemically relevant reaction between \ce{H2O+} and HD was investigated at collision energies  $E_\mathrm{coll}/k_\mathrm{B}=7$~K using a novel cryogenic single-ion experimental approach. The present work extends laboratory studies for this reaction towards the low-temperature regime of interest for interstellar chemistry. By directly counting reaction outcomes for isolated events, branching fractions for the competing H-atom and D-atom transfer channels were determined under isolated conditions. The observed preference for H-transfer observed at higher collision energies was found to persist into the cold regime revealing a significant intramolecular isotope effect which is, however, less pronounced than observed at specific higher energies. The experimental results are interpreted with the help of dynamics calculations which reveal a change of reaction mechanism from direct to indirect dynamics at the low energies of relevance here. The switch of mechanism directly impacts the product branching ratios and points to an increasing role of IVR at low energies.

We have demonstrated an experimental platform for studying cold ion-molecule reactions at the single-particle level by combining long-lived trapping of individual molecular ions, internal cooling, and non-destructive product identification. More broadly, this work establishes cold single-ion chemistry as an attractive tool for disentangling ion--molecule reaction dynamics and for testing theoretical reaction models under well-defined conditions. In the future, this platform will enable studies of state-dependent collisions of molecular ions, including measurements of elastic-to-inelastic scattering ratios relevant for buffer-gas cooling. Integration with Stark- or Zeeman-decelerated neutral beams \cite{haas17a} and of minimally perturbative state-readout techniques \cite{sinhal20a} offers pathways toward simultaneous control of collision energy and internal state, and ultimately toward full state-to-state resolution.


\section*{\label{sec:Experiment}Methods}

\subsection*{Experimental methods}

\paragraph{Experimental apparatus:} The experimental apparatus used in the present study consisted of a molecular-beam source for generating a supersonic molecular beam of water molecules and a linear radiofrequency (RF) quadrupole ion trap \cite{willitsch12a, meir19a} for reaction studies housed within a cryogenic environment, see Fig. \ref{fig:crystal_and_tickling}(a) for a schematic. The system is divided into differentially pumped regions, with pressures decreasing from $5\times10^{-7}$~mbar in the source chamber to $4\times10^{-14}$~mbar in the science chamber. The regions were separated by gate valves. The trap is thermally anchored to a cryogenic platform connected to a custom-made low-vibration cryostat with base temperatures down to 4~K. Two concentric cryogenic shields surrounding the trap are held at 5.4~K and 23~K under operating conditions providing cryopumping of the chamber and suppressing blackbody radiation at the location of the trap. To further improve cryopumping and reduce the residual gas load, 1.5 g of activated charcoal sorbent were introduced into the cryogenic assembly.

\paragraph{Preparation of \ce{Ca+} Coulomb crystals:} Each experimental cycle was initiated with the loading of \ce{Ca+} ions into the center of the trap. For that purpose, an effusive beam of neutral \ce{Ca} atoms was resonantly ionized using diode-laser radiation at 423~nm and 375~nm to generate \ce{Ca+} ions \cite{meir19a}. The \ce{Ca+} ions were Doppler laser cooled by driving the closed cycling transitions $(4s)~^2S_{1/2} \leftrightarrow (4p)~^2P_{1/2} \leftrightarrow (3d)~^2D_{3/2}$ using diode lasers at 397~nm and 866~nm \cite{willitsch12a} to form Coulomb crystals in the trap as shown in Fig. \ref{fig:crystal_and_tickling}(b)(i). The resonance fluorescence from the laser-cooled ions was collected by an objective and imaged onto an electron-multiplied charge-coupled device (EMCCD) camera.

\paragraph{Generation and trap loading of single water ions:} A pulsed supersonic beam of \ce{H2O} molecules was generated by expanding water vapor seeded into 1.5 bar of high-purity argon (99.996\%) as a carrier gas through a piezoelectric valve. The resulting beam was collimated by two skimmers with apertures of 1.0~mm and 0.5~mm before entering the trap. In the trap center, single water ions were generated by [2+1] resonance-enhanced multi-photon ionization (REMPI) via the $\tilde{C}$ electronic state of H$_2$O \cite{yang10a}. The required radiation at 248~nm was produced by a dye laser (LIOPTEC) operating with Coumarin 307 dye, pumped by the third harmonic of an Nd:YAG laser (Innolas Spotlight 1500) at 10 Hz. The output of the dye laser was frequency-doubled to obtain 1.5 mJ of the target wavelength, which was focused into the trap center by a lens with a focal length of 50~cm. Once the molecular ion was loaded, the gate valve isolating the trap chamber was closed to prevent streaming of background gas into the reaction region. 

\paragraph{Resonant motional excitation:} For a two-ion crystal composed of a $^{40}\mathrm{Ca}^+$ ion of mass $m_\text{Ca}$ and a dark ion of mass $m_d$, the motion along the longitudinal symmetry axis of the linear trap is described by two normal modes \cite{drewsen04a}. Their frequencies $f_\pm$ can be expressed in terms of the mass ratio $\mu = m_d/m_{\mathrm{Ca}}$ and the independently measured motional frequency of a single Ca$^+$ ion, $f_{z,\mathrm{Ca}}$, as \cite{james98a,leibfried03a},
\begin{equation}
f_{\pm}
=
f_{z,\mathrm{Ca}}
\sqrt{
\frac{1}{\mu}\left[(1+\mu)\pm\sqrt{1-\mu+\mu^2}\right]
}. \label{eq:freq}
\end{equation}
The lower-frequency mode $f_{-}$ corresponds to an in-phase (IP) motion, while the higher-frequency mode $f_{+}$ corresponds to an out-of-phase (OP) mode. 
In the present experiments, a weak sinusoidal voltage $V_{\mathrm{t}}\sin(\omega t)$ was applied to a trap electrode to excite the IP mode of the two-ion crystal \cite{kim10a,schmidt20a,home11a,fan21a}. When the drive frequency $f$ matched a resonance according to Eq.~\ref{eq:freq}, the amplitude of the motion of the ions in the trap increased, reducing the efficiency of laser cooling because of the Doppler effect \cite{drewsen04a} thus causing a change in the fluorescence of the laser-cooled \ce{Ca+} ion. This manifested itself as a blurring and swapping of the Ca$^+$ fluorescence in the images due to periodic modulation \cite{kim10a,home11a} (inset (i) in Fig.~\ref{fig:crystal_and_tickling}(b). By measuring the fluorescence response as a function of $f$ and comparing the observed mode frequency with the theoretical values for a two-ion crystal according to Eq.~\ref{eq:freq}, the mass of the trapped molecular ion was determined.

\subsection*{Theoretical methods}

\paragraph{Quasi-classical trajectory calculations} The quasi-classical trajectory (QCT) calculations were performed using the standard technique implemented in VENUS \cite{hu91} on our previous potential energy surface \cite{li14a}. The trajectories were initiated with a 13 Å separation between reactants (H$_2$O$^+$ and HD), and terminated when products reached a separation of 6 Å, or when the reactants were separated by 12 Å for non-reactive trajectories. The collision energy was systematically varied from 0.02 to 20 kcal/mol (0.9 meV to 0.9 eV). At the specified collision energy, the initial vibrational and rotational states for both reactants were kept in their ground states. The maximal impact parameter ($b_{max}$) was determined at each collision energy using small batches of trajectories with trial values and ranged from 12 Å at 0.02 kcal/mol to 6.0 Å at 10.0 kcal/mol. Energy conservation was carefully maintained for all trajectories, with a variation of less than 0.04 kcal/mol, by setting the propagation time step to 0.02 fs. If a trajectory took longer than 4 ps to propagate, it was discarded. Statistical accuracy was ensured by running batches of 20,000 trajectories at each collision energy. The reaction probabilities for the H-atom transfer and D-atom transfer reaction,$P_r^H$ and $P_r^D$, were given by the ratio between the number of reactive trajectories ($N_r^H$ and $N_r^D$) and the total number of trajectories ($N_{total}$), respectively, and their branching ratio is $P_r^H / P_r^D = N_r^H / N_r^D $.

\paragraph{Criteria for classifying direct and trapped trajectories} A trajectory was considered to be residing in the pre-reaction well if the following three conditions are met simultaneously: (1) the distance between the oxygen atom and the center of mass of the incident HD lies between $1.6$ and $4.0$~Å, indicating that the system is within the short range interaction region but not yet in the tight product complex; (2) the internuclear distance of the incident HD is less than $1.1$~Å, ensuring that the HD bond remains largely intact and has not undergone significant elongation or rupture; and (3) the distance from the oxygen atom to the third nearest H/D atom is greater than $1.35$~Å, which excludes configurations where a new O–H/D bond has already been formed. A trajectory was classified as “trapped” if these three conditions were continuously satisfied for more than $200$~fs, corresponding to a sufficiently long residence time in the well to allow for efficient IVR (see Sec. \ref{sec:Complex_lifetime} for details). Conversely, if the continuous well‑residence time is $200$~fs or less, the trajectory is considered “direct”, implying that the system passes through the interaction region and forms products without substantial trapping.

\section*{Acknowledgements}

We thank Georg Holderied, Dominik Baumgartner, Anatoly Johnson, Grischa Martin and Philipp Knöpfel for technical support. This work was supported financially by the Swiss National Science Foundation (grant nr. TMAG-2\_209193) and the University of Basel. P.P. thanks the G. H. Endress foundation for financial support. A.L. acknowledges the support of the National Natural Science Foundation of China (Grant No.22273072). H.G. is supported by Department of Energy (Grant No. DE-SC0015997).

\printbibliography

@string{apj = {{Astrophys.~J.}}}

@string{cpl = {{Chem. Phys. Lett.}}}

@string{fd = {{Faraday Discuss.}}}

@string{ijms = {{Int. J.~Mass Spectrom.}}}

@string{irpc = {{Int. Rev. Phys. Chem.}}}

@string{jcc = {{J.~Comput. Chem.}}}

@string{jcp = {{J.~Chem. Phys.}}}

@string{jpca = {{J.~Phys. Chem.~A}}}

@string{pccp = {{Phys. Chem. Chem. Phys.}}}

@string{prl = {{Phys. Rev. Lett.}}}

@string{rmp = {{Rev. Mod. Phys.}}}

@string{science = {{Science}}}

@string{sad = {{Sci. Adv.}}}

@string{mnrasl = {{Mon.~Not.~R.~Astron.~Soc.:Lett.}}}

@string{cern = {{CERN Yellow Rep.: Sch. Proc.}}}

@string{jpcl = {{J.~Phys. Chem. Lett.}}}

@string{Wiley = "Wiley"}

@ARTICLE{snow08a,
  title     = "Ion chemistry in the interstellar medium",
  author    = "Snow, Theodore P and Bierbaum, Veronica M",
  journal   = "Annu. Rev. Anal. Chem.",
  publisher = "Annual Reviews",
  volume    =  1,
  number    =  1,
  pages     = "229--259",
  year      =  2008,
  language  = "english"
}

@ARTICLE{larsson12a,
  title     = "Ion chemistry in space",
  author    = "Larsson, M and Geppert, W D and Nyman, G",
  journal   = "Rep. Prog. Phys.",
  publisher = "IOP Publishing",
  volume    =  75,
  number    =  6,
  pages     = "066901",
  month     =  jun,
  year      =  2012,
  language  = "english"
}

@article{kumar18a,
  title={Low temperature rates for key steps of interstellar gas-phase water formation},
  author={Kumar, Sunil S and Grussie, Florian and Suleimanov, Yury V and Guo, Hua and Kreckel, Holger},
  journal=sad,
  volume={4},
  number={6},
  pages={eaar3417},
  year={2018},
  publisher={American Association for the Advancement of Science}
}

@article{tran18a,
  title={Formation of $\mathrm{H_2O^+}$ and $\mathrm{H_3O^+}$ Cations in Reactions of $\mathrm{OH^+}$ and $\mathrm{H_2O^+}$ with $\mathrm{H_2}$: Experimental Studies of the Reaction Rate Coefficients from $T = 15$ to $300\,\mathrm{K}$},
  author={Tran, Thuy Dung and Rednyk, Serhiy and Kovalenko, Artem and Rou{\v{c}}ka, {\v{S}}t{\v{e}}p{\'a}n and Dohnal, Petr and Pla{\v{s}}il, Radek and Gerlich, Dieter and Glos{\'\i}k, Juraj},
  journal=apj,
  volume={854},
  number={1},
  pages={25},
  year={2018},
  publisher={IOP Publishing}
}

@article{ard14,
  title={Experimental and Theoretical Kinetics for the $\mathrm{H_2O^+ + H_2/D_2 \rightarrow H_3O^+/H_2DO^+ + H/D}$ Reactions: Observation of the Rotational Effect in the Temperature Dependence},
  author={Ard, Shaun G and Li, Anyang and Martinez Jr, Oscar and Shuman, Nicholas S and Viggiano, Albert A and Guo, Hua},
  journal=jpca,
  volume={118},
  number={49},
  pages={11485--11489},
  year={2014},
  publisher={ACS Publications}
}

@article{song17a,
  title={Competition between the H- and D-Atom Transfer Channels in the $\mathrm{H_2O^+ + HD}$ Reaction: Reduced-Dimensional Quantum and Quasi-Classical Studies},
  author={Song, Hongwei and Li, Anyang and Yang, Minghui and Guo, Hua},
  journal=pccp,
  volume={19},
  number={26},
  pages={17396--17403},
  year={2017},
  publisher={Royal Society of Chemistry}
}

@article{xu17a,
  title={Isotopic and quantum-rovibrational-state effects for the ion--molecule reaction in the collision energy range of 0.03--10.00 eV},
  author={Xu, Yuntao and Xiong, Bo and Chang, Yih Chung and Ng, C. Y.},
  journal=pccp,
  volume={19},
  number={13},
  pages={8694--8705},
  year={2017},
  publisher={Royal Society of Chemistry}
}

@article{xu12a,
	author = {Y. Xu and B. Xiong and Y. C. Chang and C. Y. Ng},
	doi = {10.1063/1.4773099},
	journal = jcp,
	title = {Communication: Rovibrationally selected absolute total cross sections for the reaction $\mathrm{H_2O^+}(X\,^2B_1;
v_1^+v_2^+v_3^+ = 000;
N^+_{K_a^+K_c^+}) + \mathrm{D_2}$:
Observation of the rotational enhancement effect},
	pages = {241101},
	volume = {137},
	year = {2012}
}

@article{balashev19a,
    author = {Balashev, S A and Kosenko, D N},
    title = {HD/H$_2$ ratio in the diffuse interstellar medium},
    journal = mnrasl,
    volume = {492},
    number = {1},
    pages = {L45-L49},
    year = {2020},
    publisher={Oxford University Press}
}

@ARTICLE{wild23a,
  title    = "Tunnelling measured in a very slow ion-molecule reaction",
  author   = "Wild, Robert and N{\"o}tzold, Markus and Simpson, Malcolm and
              Tran, Thuy Dung and Wester, Roland",
  journal  = "Nature",
  volume   =  615,
  number   =  7952,
  pages    = "425--429",
  month    =  mar,
  year     =  2023,
  language = "english"
}

@ARTICLE{hugo09a,
  title     = "$\mathrm{H_3^+ + H_2}$ isotopic system at low temperatures: microcanonical
               model and experimental study",
  author    = "Hugo, Edouard and Asvany, Oskar and Schlemmer, Stephan",
  journal   = jcp,
  publisher = "AIP Publishing",
  volume    =  130,
  number    =  16,
  pages     = "164302",
  month     =  apr,
  year      =  2009,
  language  = "en"
}

@article{wang26a,
  author  = {Wang, Chen and Gan, Wen-Ting and Peng, Wen-Cui and Tong, Xin},
  title   = {Cold Chemistry Probes Isotope Effects in the {Be$^+$} + {HD} Reaction},
  journal = jpcl,
  year    = {2026},
  volume  = {17},
  number  = {29},
  pages   = {8173--8178},
  doi     = {10.1021/acs.jpclett.6c01524},
  url     = {https://doi.org/10.1021/acs.jpclett.6c01524}
}

@article{willitsch12a,
	author = {Willitsch, S.},
	journal = irpc,
	title = {Coulomb-crystallised molecular ions in traps: methods, applications, prospects},
	pages = {175-199},
	volume = {31},
	year = {2012}}

@article{meir19a,
	author = {Meir, Ziv and Hegi, Gregor and Najafian, Kaveh and Sinhal, Mudit and Willitsch, Stefan},
	doi = {10.1039/C8FD00195B},
	journal = fd,
	title = {State-selective coherent motional excitation as a new approach for the manipulation{,} spectroscopy and state-to-state chemistry of single molecular ions},
	pages = {561-583},
	volume = {217},
	year = {2019}}

@article{uselman07a,
author = {Brady W. Uselman and Jason M. Boyle and Scott L. Anderson},
title = {Multiphoton ionization vibrational state selection of $\mathrm{H_2O^+}$, $\mathrm{D_2O^+}$ and $\mathrm{HOD^+}$},
journal = cpl,
volume = {440},
number = {4},
pages = {171-175},
year = {2007},
issn = {0009-2614},
doi = {10.1016/j.cplett.2007.04.030},
url = {https://www.sciencedirect.com/science/article/pii/S0009261407004770},
}

@article{willitsch05b,
	author = {Willitsch, S. and Merkt, F.},
	journal = ijms,
	title = {Rovibronic photoionization dynamics of asymmetric-top molecules},
	pages = {14-25},
	volume = {245},
	year = {2005}}

@article{haas17a,
	author = {Haas, Dominik and Scherb, Sebastian and Zhang, Dongdong and Willitsch, Stefan},
	journal = {EPJ Tech. Instrum.},
	title = {Optimizing the Density of {{Stark}} Decelerated Radicals at Low Final Velocities: A Tutorial Review},
	pages = {6},
	volume = {4},
	year = {2017}}

@article{sinhal20a,
	author = {Sinhal, Mudit and Meir, Ziv and Najafian, Kaveh and Hegi, Gregor and Willitsch, Stefan},
	doi = {10.1126/science.aaz9837},
	journal = Science,
	title = {Quantum-non demolition state detection and spectroscopy of single trapped molecules},
	pages = {1213-1218},
	volume = {367},
	year = {2020}}

@article{yang10a,
	author = {C. H. Yang and G. Sarma and J. J. ter Meulen and D. H. Parker and C. M. Western},
	doi = {10.1039/c0cp00946f},
	journal = pccp,
	title = {{REMPI} spectroscopy and predissociation of the {$\tilde{C} {}^{1}B_1 (v = 0)$} rotational levels of {H$_2$O, HOD and D$_2$O}},
	pages = {13983-13991},
	volume = {12},
	year = {2010}}

@article{drewsen04a,
	author = {Drewsen, M. and Mortensen, A. and Martinussen, R. and Staanum, P. and S{\o}rensen, J. L.},
	journal = PRL,
	title = {Nondestructive identification of cold and extremely localized single molecular ions},
	pages = {243201},
	volume = {93},
	year = {2004}}

@article{james98a,
	author = {James, D. F. V.},
	journal = {Appl. Phys. B},
	title = {Quantum dynamics of cold trapped ions with application to quantum computation},
	pages = {181-190},
	volume = {66},
	year = {1998}}

@article{leibfried03a,
	author = {Leibfried, D and Blatt, R and Monroe, C and Wineland, D},
	journal = rmp,
	title = {{Quantum dynamics of single trapped ions}},
	pages = {281-324},
	volume = {75},
	year = {2003}}

@article{kim10a,
	author = {Kim, K. and Chang, M.-S. and Korenblit, S. and Islam, R. and Edwards, E. E. and Freericks, J. K. and Lin, G.-D. and Duan, L.-M. and Monroe, C.},
	journal = {Nature},
	title = {Quantum simulation of frustrated Ising spins with trapped ions},
	pages = {590-593},
	volume = {465},
	year = {2010}}

@article{schmidt20a,
	author = {J. Schmidt and T. Louvradoux and J. Heinrich and N. Sillitoe and M. Simpson and J.-P. Karr and L. Hilico},
	journal = {Phys. Rev. Appl.},
	title = {Trapping, Cooling, and Photodissociation Analysis of State-Selected H$_2^+$ Ions Produced by (3 + 1) Multiphoton Ionization},
	pages = {024053},
	volume = {14},
	year = {2020}}

@article{home11a,
	author = {Home, J. P. and Hanneke, D. and Jost, J. D. and Leibfried, D. and Wineland, D. J.},
	journal = {New J. Phys.},
	title = {{Normal modes of trapped ions in the presence of anharmonic trap potentials}},
	pages = {073026},
	volume = {13},
	year = {2011}}

@ARTICLE{fan21a,
  title     = {Optical Mass Spectrometry of Cold $\mathrm{RaOH^+}$ and $\mathrm{RaOCH_3^+}$},
  author    = "Fan, M and Holliman, C A and Shi, X and Zhang, H and Straus, M W
               and Li, X and Buechele, S W and Jayich, A M",
  journal   = prl,
  publisher = "American Physical Society (APS)",
  volume    =  126,
  number    =  2,
  pages     = "023002",
  month     =  jan,
  year      =  2021,
  copyright = "https://link.aps.org/licenses/aps-default-license",
  language  = "en"
}

@article{hu91,
  title={Vectorization of the general Monte Carlo classical trajectory program VENUS},
  author={Hu, Xiche and Hase, William L and Pirraglia, Tony},
  journal=jcc,
  volume={12},
  number={8},
  pages={1014--1024},
  year={1991},
  publisher={Wiley Online Library}
}

@article{li14a,
	author = {Li, A. and Guo, H.},
	journal = jcp,
	title = {A nine-dimensional ab initio global potential energy surface for the $\mathrm{H_2O^+ + H_2 \rightarrow H_3^+ + H}$ reaction},
	pages = {224313},
	volume = {140},
	year = {2014}}

@techreport{chiggiato14a,
  author       = {Chiggiato, P.},
  title        = {Vacuum Technology for Ion Sources},
  type         = {CERN Yellow Report},
  number       = {CERN-2013-007},
  institution  = {CERN},
  organization = {CERN},
  address      = {Geneva, Switzerland},
  pages        = {463--502},
  year         = {2013}
}

@techreport{Roder1973a,
  author      = {Roder, H. M. and Childs, G. E. and
                 McCarty, R. D. and Angerhofer, P. E.},
  title       = {Survey of the Properties of the Hydrogen
                 Isotopes Below Their Critical Temperatures},
  institution = {National Bureau of Standards},
  type        = {NBS Technical Note},
  number      = {641},
  address     = {Washington, DC},
  year        = {1973}
}

@article{Scannapiego17a,
  author  = {Scannapiego, M. and Day, C.},
  title   = {Experimental investigation on charcoal
             adsorption for cryogenic pump application},
  journal = {IOP Conf. Ser.: Mater. Sci. Eng.},
  volume  = {278},
  number  = {1},
  pages   = {012160},
  year    = {2017}
}

@article{rouse15a,
	author = {Rouse, I and Willitsch, S},
	journal = {Phys. Rev. A},
	title = {{Superstatistical velocity distributions of cold trapped ions in molecular-dynamics simulations}},
	pages = {053420},
	volume = {92},
	year = {2015}}

\clearpage
\newpage
\renewcommand{\thesection}{S\arabic{section}}
\setcounter{section}{0}
\renewcommand{\thefigure}{S\arabic{figure}}
\setcounter{figure}{0}
\renewcommand{\thetable}{S\arabic{table}}
\setcounter{table}{0}
\setcounter{page}{1}

\begin{center}
\noindent {\LARGE\bf Supporting Information}

    \vspace{0.5cm}

    {\LARGE\bfseries
    Single-molecule cold chemistry: Isotopic branching ratios of the \ce{H2O+} + HD reaction at cryogenic temperatures
    }

\begin{center}

{\large
Nanditha Sunil Kumar\textsuperscript{1*},
Prerna Paliwal\textsuperscript{1*},
Mikhail Popov\textsuperscript{1},
Xueyi Li\textsuperscript{2}
Anyang Li\textsuperscript{2},
Hua Guo\textsuperscript{3}, and
Stefan Willitsch\textsuperscript{1}
\par}

\vspace{1 cm}

\textsuperscript{1}Department of Chemistry,
University of Basel, Klingelbergstrasse 80,
4056 Basel, Switzerland

\par
\textsuperscript{2}Key Laboratory of Synthetic
and Natural Functional Molecule of the Ministry
of Education, College of Chemistry and Materials
Science, Northwest University, 710127 Xi'an, China

\par
\textsuperscript{3}Department of Chemistry and
Chemical Biology, Center for Computational
Chemistry, University of New Mexico,
Albuquerque, New Mexico 87131, USA

\vspace{1 cm}

Email: liay@nwu.edu.cn (Anyang Li),
hguo@unm.edu (Hua Guo),
stefan.willitsch@unibas.ch (Stefan Willitsch)

\end{center}

\end{center}

\vspace{1 cm}

\clearpage

\section{Resonance frequencies of two-ion strings}
\label{sec:frequencies}

\begin{table}[!h]
\caption{Resonance frequencies at the present trapping conditions for the excitation of the axial in-phase (IP) mode of a two-ion string consisting of single Ca$^+$ and molecular ions of relevance in the present study.}
\label{tab:frequencies_table}
\centering
\begin{tabular}{l l l}
\hline
Ion pair & Mode & Frequency (kHz) \\
\hline
 \ce{Ca+}--\ce{H2O+}& Axial IP  & 531.9 kHz \\
 \ce{Ca+}--\ce{H3O+}& Axial IP & 529.2 kHz \\
 \ce{Ca+}--\ce{H2DO+}& Axial IP & 526.4 kHz \\
\hline
\end{tabular}
\end{table}


\section{Partial pressures of HD and \texorpdfstring{H$_2$}{H2}}
\label{sec:pressures}

Partial pressures were estimated from the measured reactive-loss lifetimes at
$T=5.4~\mathrm{K}$ using
\begin{equation}
p_i=\frac{f_i k_B T}{\tau\sum_j f_j k_{L,j}},
\end{equation}
where $f_i$ is the gas fraction ($f_i=1$ for the background H$_2$ measurement) and $k_{L,j}$ is the Langevin rate coefficient for the gas species $j$ in the mixture.

\begin{table}[h]
\centering
\caption{Estimated HD and H$_2$ partial pressures assuming a fraction of $f_{\mathrm{HD}}=0.97$ and $f_{\mathrm{H_2}}=0.03$ in the HD gas sample used.}
\label{tab:pressure_meas}
\begin{tabular}{lccc}
\hline
Condition & $k_{L,j}$ (cm$^3$\,s$^{-1}$) & Lifetime & Pressure (mbar) \\
\hline
HD after dosing
& $1.29\times10^{-9}$ & $98\pm44$ s
& $(5.7\pm2.5)\times10^{-12}$ \\
H$_2$ after dosing
& $1.54\times10^{-9}$ & $98\pm44$ s
& $(1.8\pm0.8)\times10^{-13}$ \\
Background H$_2$
& $1.54\times10^{-9}$ & $4.5\pm1.0$ h
& ($3.0\pm0.7)\times10^{-14}$ \\
\hline
\end{tabular}
\end{table}


\section{Characterization of background reactions}
\label{sec:bgreactions}

\begin{figure}
  \centering
  \includegraphics[width=\linewidth]{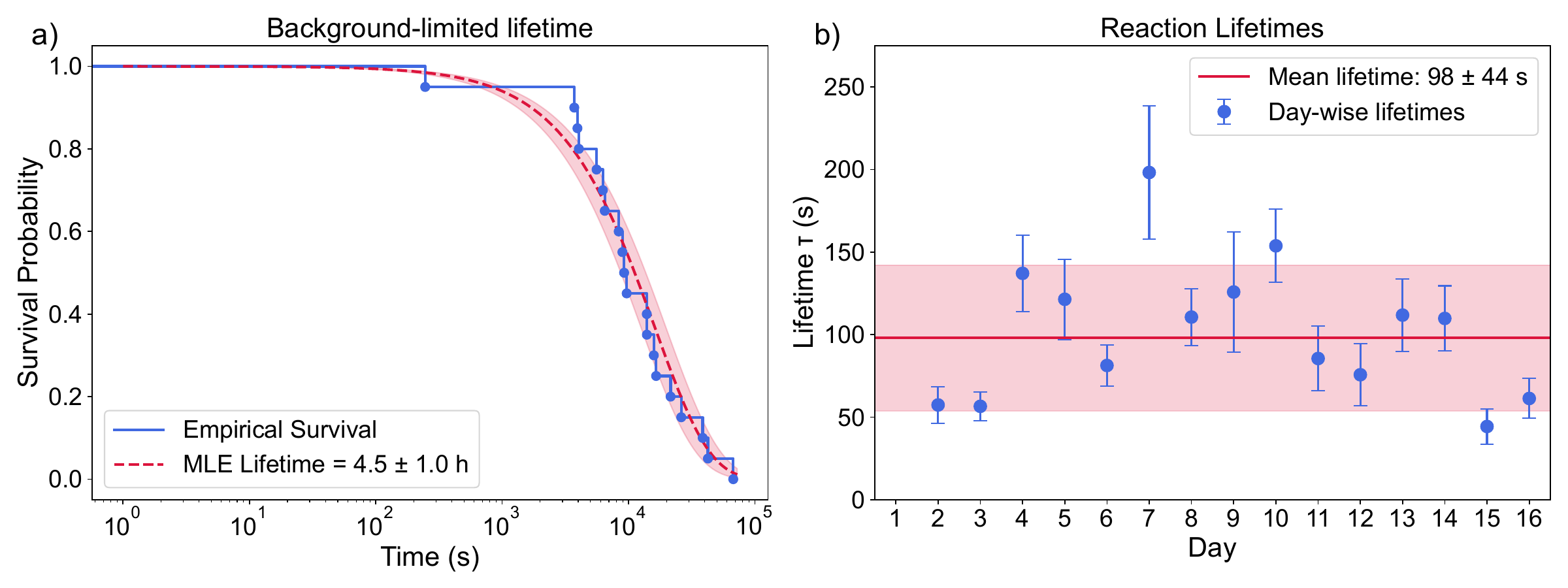}
  \caption{Lifetimes of trapped \ce{H2O+} limited by reactive collisions. \textbf{(a)} Survival of trapped \ce{H2O+} ions limited by reactions with background H$_2$ gas. The empirical survival probability as a function of time $t$ (blue points/line), defined as the fraction of ions with lifetime $\ge t$, was fit with a single-exponential model using maximum-likelihood estimation (MLE), yielding a lifetime $\tau_{\mathrm{bg}} = 4.5 \pm 1.0~\mathrm{h}$ (red dashed line; shaded band indicates the fit uncertainty). \textbf{(b)}~Lifetime of trapped H$_2$O$^+$ ions limited by reactive loss in the presence of HD. The lifetimes determined on each measurement day (blue points with error bars) were obtained from a maximum-likelihood estimation (MLE) analysis performed independently for each day. The mean reactive-loss lifetime, $\tau_{\mathrm{HD}} = 98 \pm 44~\mathrm{s}$, was calculated as the average of the daily values.} 
  \label{fig:Lifetimes_H2_and_HD}
\end{figure}

The mean lifetime $\tau_\mathrm{bg}$ of the trapped water ions limited by reactions with background gas in the vacuum chamber was determined from 20 independent reaction events under identical conditions. Ion loss was modeled as a Poissonian process with a time-independent loss rate. Accordingly, the survival probability, defined as the fraction of ions remaining after a storage time $t$ was described by,
\begin{equation}
S(t)=\exp\!\left(-\frac{t}{\tau_{\mathrm{bg}}}\right),
\end{equation}
where $\tau_{\mathrm{bg}}$ is the background-gas-limited trap lifetime. 
$\tau_{\mathrm{bg}}$ was determined using a maximum-likelihood estimation (MLE) applied directly to the set of measured lifetimes, yielding $\tau_{\mathrm{bg}} = 4.5(1.0)~\mathrm{h}$ (Fig. \ref{fig:Lifetimes_H2_and_HD}a), with the quoted uncertainty given by the statistical error of the MLE fit. Assuming Langevin kinetics and the ideal gas law, the background $\mathrm{H_2}$ partial pressure was estimated from the measured reaction time $\tau_{\mathrm{bg}}$ as
\begin{equation}
P_{\mathrm{H_2}} = \frac{k_B T}{k_{L, \text{H}_2}\,\tau_{\mathrm{bg}}},
\end{equation}
where $k_L$ the Langevin rate coefficient (Table~\ref{tab:pressure_meas}). The resulting value was found to be $P_{\mathrm{H_2}}=3.0(0.7)\times10^{-14}\,~\mathrm{mbar}$.


\section{Correction of observed branching ratio for sample impurities}
\label{sec:branalysis}

\subsection{Correction for \texorpdfstring{H$_2$}{H2} contamination for product branching ratios}
For gas fractions $f_{\mathrm{HD}}+f_{\mathrm{H_2}}=1$ and bimolecular rate constants $k_{\mathrm{HD}}$ and $k_{\mathrm{H_2}}$, the fraction of reactive events due to H$_2$ is
\begin{equation}
\alpha =
\frac{f_{\mathrm{H_2}}k_{\mathrm{H_2}}}
{f_{\mathrm{HD}}k_{\mathrm{HD}}+f_{\mathrm{H_2}}k_{\mathrm{H_2}}}.
\label{eq:alpha}
\end{equation}

Assuming H$_2$ produces only $\mathrm{H_3O^+}$ and HD produces both channels, the observed product ratio $r_{\mathrm{obs}}=N(\mathrm{H_3O^+})/N(\mathrm{H_2DO^+})$ must consequently be corrected. We define $A_{\mathrm{tot}} = N(\ce{H3O+})$ and $ B_{\mathrm{tot}} = N(\ce{H2DO+})$ as the totally measured
counts of \ce{H3O+} and \ce{H2DO+}, respectively, and $r_{\mathrm{obs}}={A_{\mathrm{tot}}}/{B_{\mathrm{tot}}}$. The total number of
detected events is $N_{\mathrm{tot}} = A_{\mathrm{tot}} + B_{\mathrm{tot}}$. Thus, the corrected product branching ratio is obtained as,

\begin{align}
r_{\mathrm{corr}}
&=
\frac{A_{\mathrm{tot}}-\alpha N_{\mathrm{tot}}}
     {B_{\mathrm{tot}}}
\\
&=
\frac{A_{\mathrm{tot}}
-\alpha\left(A_{\mathrm{tot}}+B_{\mathrm{tot}}\right)}
{B_{\mathrm{tot}}}
\\
&= r_{\mathrm{obs}}(1-\alpha)-\alpha. 
\label{eq:correction}
\end{align}

\begin{figure}[htbp]
  \centering
  \includegraphics[width=0.6\linewidth]{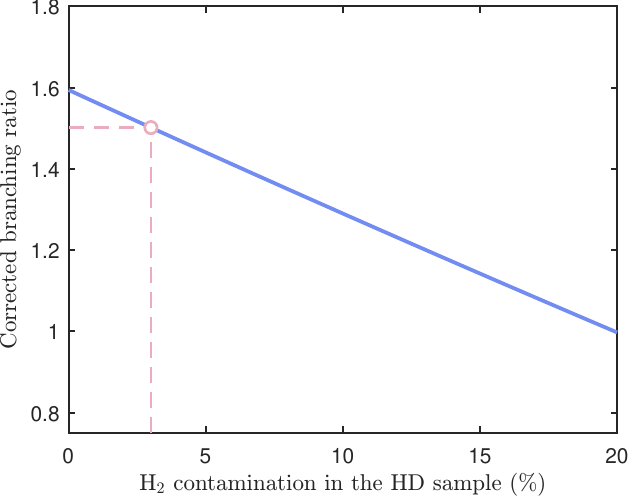}
    \caption{Corrected $\mathrm{H_3O^+/H_2DO^+}$ product ratio as a function of the assumed $\mathrm{H_2}$ contamination in the HD sample. For the present sample contamination of 3\% H$_2$, a correct product ratio of 1.50 is obtained (dashed lines).}
  \label{fig:Branching_ratio_vs_H2_contamination}
\end{figure}

\begin{figure}[htbp]
  \centering
  \includegraphics[width=0.6\linewidth]{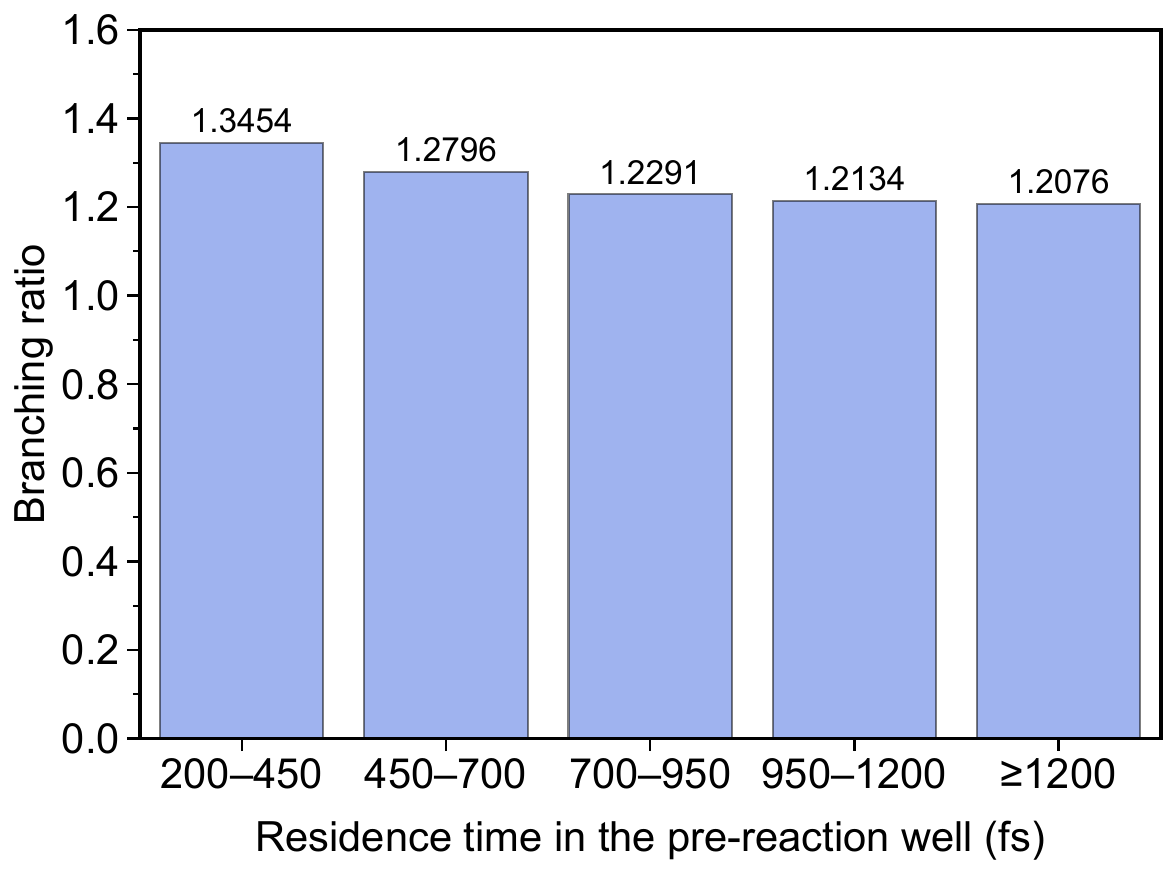}
  \caption{Branching ratio as a function of the continuous residence time in the pre-reaction well for trapped trajectories at a collision energy of 0.02~kcal/mol (10~K).}
  \label{fig:complex_lifetime}
\end{figure}

The correction to the product ratio according to eq.~(\ref{eq:correction}) is shown in Fig.~\ref{fig:Branching_ratio_vs_H2_contamination}. The correction assumes capture-limited Langevin reaction rates with the contaminant contribution weighted by $k_{\mathrm{L,H_2}}/k_{\mathrm{L_{HD}}} = 1.19$. Using the measured yields of 271 $\mathrm{H_3O^+}$ and 170 $\mathrm{H_2DO^+}$ events, the expected $\mathrm{H_3O^+}$ contribution from contaminant $\mathrm{H_2}$ is subtracted to obtain the intrinsic HD branching ratio. For an assumed $\mathrm{H_2}$ contamination of 3\%, the corrected ratio is $\mathrm{H_3O^+/H_2DO^+} = 1.50$.

\subsection{Discussion of possible processes affecting the gas composition in the trap volume}
The gas composition in the trap volume can potentially vary from the nominal composition of the sample and be modified by several processes in the cryogenic environment that are not independently constrained. Under conditions in which the reactant gas effuses into the reaction volume through small apertures, the partial pressure of the gas near the ion trap may not directly reflect the original mixture at room temperature but depends on the conductance through the apertures approximately scaling as 1/$\sqrt{m}$ where $m$ is the molecular mass \cite{chiggiato14a}. However, on the long time scales of the present experiment, we expect mass-dependent effusion effects to be negligible. In addition, at temperatures below $\sim$10~K, isotope-dependent condensation on the trap walls may introduce further fractionation \cite{Roder1973a}. At the operating temperature of the trap, cryosorption on the charcoal provides an additional selective and time-dependent loss channel for H$_2$ and HD \cite{Scannapiego17a}. Because the sticking coefficients and residence times of these species on charcoal are not known at the relevant temperatures, and may also depend on surface coverage of absorbent, the ramifications of these effects cannot be quantified at present. Thus, the actual H$_2$ partial pressure near the trap may be higher than the nominally expected value which may imply an overestimation of the present product branching ratio. 


\section{Collision energy calculation}
\label{sec:ecoll}
Assuming isotropic collisions between \ce{H2O+} and HD, the mean collision energy is given by
\begin{equation}
\langle E_{\mathrm{coll}}\rangle
=
\frac{1}{2}\mu
\left(
\langle v_{\ce{H2O+}}^2\rangle
+
\langle v_{\mathrm{HD}}^2\rangle
\right),
\label{eq:Ecoll_mean}
\end{equation}
where $\mu=m_{\ce{H2O+}}m_{\mathrm{HD}}/(m_{\ce{H2O+}}+m_{\mathrm{HD}})$ is the reduced mass.

HD is assumed to thermalize to the copper housing at
$T_{\mathrm{HD}}=5.4~\mathrm{K}$. For a Maxwell--Boltzmann distribution, this corresponds to
$\langle v_{\mathrm{HD}}^2\rangle=3k_BT_{\mathrm{HD}}/m_{\mathrm{HD}}
=4.45\times10^4~\mathrm{m^2\,s^{-2}}$.

The \ce{Ca+}--\ce{H2O+} crystal is driven at the axial in-phase mode frequency
$f=531.9~\mathrm{kHz}$. Molecular dynamics simulations \cite{rouse15a} give the \ce{H2O+} kinetic energy of approximately $0.04~\mathrm{meV}$ corresponding to
$\langle v_{\ce{H2O+}}^2\rangle=4.25\times10^2~\mathrm{m^2\,s^{-2}}$. Additionally, residual RF fields and coupling between motional modes may contribute to micromotion, which we estimate to introduce an additional collision energy contribution of $\approx 0.01~\mathrm{meV}$ .

Using $\mu=2.57~\mathrm{amu}$ in Eq.~\eqref{eq:Ecoll_mean}, we obtain
\begin{equation}
\langle E_{\mathrm{coll}}\rangle
= 0.6~\mathrm{meV}
= 7~\mathrm{K}.
\end{equation}
The collision energy is therefore governed by the thermal motion of HD.


\section{Product branching ratio as a function complex lifetime}
\label{sec:Complex_lifetime}

\end{document}